\documentstyle[11pt,newpasp,twoside,epsf]{article}
\def\edcomment#1{\iffalse\marginpar{\raggedright\sl#1\/}\else\relax\fi}
\reversemarginpar

\begin{document}
\title{The Local Group}
 \author{Eva K.\ Grebel$^{1,2,3}$}
\affil{$^{1}$University of Washington, Department of Astronomy, Box 351580,
Seattle, WA 98195-1580, USA
}
\affil{$^2$Hubble Fellow}
\affil{$^3$Max-Planck-Institut f\"ur Astronomie, K\"onigstuhl 17, 
D-69117 Heidelberg, Germany}

\begin{abstract}
Local Group galaxies such as the Milky Way, the Magellanic Clouds and M31
are being used by a number of international collaborations to search for
microlensing events. Type and number of detections place constraints on dark
matter and the stellar populations within and along the line of sight to
these galaxies.  In this review I briefly discuss 
the stellar populations, evolutionary histories, and other properties
of different types of Local Group galaxies as well as 
constraints on the dark matter content of these galaxies. 
Particular emphasis is placed on the dwarf
companions of the spiral galaxies in the Local Group.
\end{abstract}

\section{Introduction}

The ``Local Group'' is the small group of galaxies around the Milky Way and
M31.  The size of the Local Group is not well known, and its galaxy
census is incomplete for low-surface-brightness galaxies.   Recent
studies suggest that the radius of the zero-velocity surface of the Local 
Group is $\sim 1.2$ Mpc (Courteau \& van den Bergh 1999) when a spherical
potential is assumed.  Within this radius 35 galaxies have been detected 
(see Grebel 2000a for a list).  
Since information about orbits is lacking it is unknown which ones of the 
more distant galaxies within and just outside of the adopted Local Group 
boundaries are actually bound to
the Local Group.  Many faint Local Group galaxies were only discovered
in recent years, and searches are continuing.  Hierarchical cold dark 
matter (CDM) models predict 
about 10 times more dark matter halos than the number of known Local Group 
satellites (e.g., Klypin et al.\ 1999).  Compact high-velocity clouds 
(Braun \& Burton 1999), which appear to be dark-matter-dominated 
with total estimated masses of a few $10^8$ M$_{\odot}$ may be 
good candidates for the ``missing'' satellites.

The Local Group comprises galaxies with a variety of different morphological
types, a range of masses, ages, and metallicities, and differing degrees of 
isolation.  Their proximity makes these galaxies ideal targets for detailed 
studies
of their star formation histories from their resolved stellar populations
and of galaxy evolution in general.  Furthermore, Local Group galaxies 
provide a convenient set of targets for studies of the nature of dark
matter.  Several Local Group reviews have appeared
in the past few years, including Mateo (1998), Grebel (1997, 1999, 2000a),
and the very detailed recent reviews by van den Bergh (1999, 2000).
Reviews dealing with dark matter in Local Group dwarf spheroidals include
Mateo (1997) and Olszewski (1998). 

\begin{figure}[ht]
\plotone{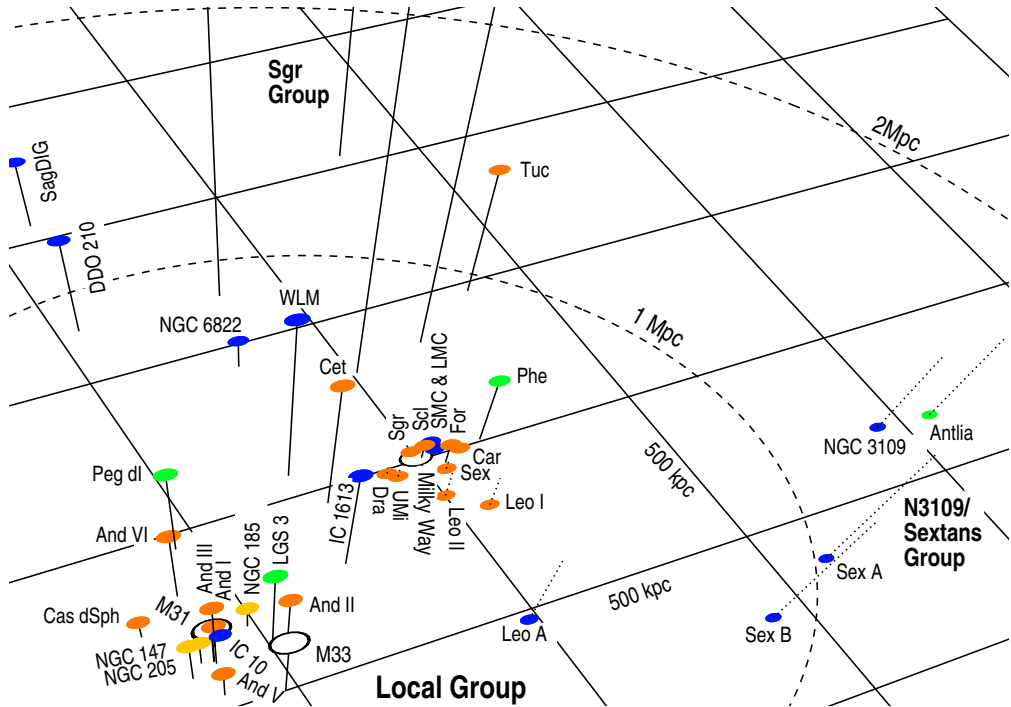}
\caption{A scaled 3-D representation of the Local Group (LG).
The dashed ellipsoid marks a radius
of 1 Mpc around the LG barycenter (assumed to be at 462 kpc toward
$l=121.7$ and $b=-21.3$ following Courteau \& van den Bergh 1999).
Distances of galaxies from the the arbitrarily chosen plane 
through the Milky Way are indicated by solid lines (above the plane)
and dotted lines (below). 
Morphological segregation is evident:  The dEs and gas-deficient
dSphs (light symbols) are closely concentrated around the large spirals
(open symbols). DSph/dIrr transition types
(e.g., Pegasus, LGS\,3, Phoenix) tend to be somewhat
more distant.  Most dIrrs (dark symbols)
are fairly isolated and located at larger distances.  Also indicated
are the locations of two nearby groups.}
\end{figure}

\section{Local Group galaxy content and distribution}

The most massive and most luminous Local Group galaxies are the two
spirals Milky Way and M31 ($\approx 10^{12} M_{\odot}$, M$_V \la -21$ mag).
The third, less luminous and less massive Local Group spiral M33 does 
not have any known companions and belongs to the M31 subsystem.  
About two thirds of the Local Group galaxies are found within 
300 kpc of the two spirals.  The majority of these close companions are 
dwarf spheroidal and dwarf elliptical galaxies.  
The ensemble of dwarf irregular galaxies, on the other hand, shows little 
concentration toward the two large spirals (although the two most massive
Local Group irregulars, the Large and the Small Magellanic Cloud (LMC and
SMC), are
close neighbors of the Milky Way and interact with it as well as with each 
other).  This correlation between
morphological type and distance from massive galaxies is also known
as morphological segregation and may be to some extent a consequence of
evolutionary effects.

Whether a galaxy should be considered a dwarf galaxy is somewhat arbitrary,
and different authors use different criteria.
For the purpose of this review all galaxies with
$M_V>-18$ mag will be considered dwarf galaxies, which results in 31 dwarfs,
excluding only the three spirals and the 
LMC.
We distinguish the following basic types of dwarf galaxies in the Local Group:  
\begin{itemize}
\item Dwarf irregulars (dIrrs)
with M$_V \ga -18$ mag, $\mu_V \la 23$ mag arcsec$^{-2}$, R $\la 5$ kpc, 
$M_{\rm H\,{\sc i}}
\la 10^9 M_{\odot}$, and $M_{tot} \la 10^{10} M_{\odot}$.  DIrrs are
irregular in their optical appearance, gas-rich, and show 
current or recent star formation.  Several of the dIrrs contain globular
or open clusters.
\item Dwarf ellipticals (dEs) 
with M$_V \ga -17$ mag, $\mu_V \la 21$ mag arcsec$^{-2}$, R $\la 4$ kpc, 
$M_{\rm H\,{\sc i}}
\la 10^8 M_{\odot}$, and $M_{tot} \la 10^9 M_{\odot}$.  DEs look 
globular-cluster-like in their visual appearance with a 
pronounced central concentration.  All dEs are companions of M31.  
Two of the four dEs (M32, NGC\,205) are nucleated.  
M32, a dE very close to M31, has a central black hole and follows the
same scaling relations as large elliptical galaxies, whereas the other dEs 
resemble dSphs and are therefore called spheroidals by van den Bergh (1999,
2000).  All dEs except for M32 contain globular
clusters. 
\item Dwarf spheroidals (dSphs) with
M$_V \ga -14$ mag, $\mu_V \ga 22$ mag arcsec$^{-2}$, R $\la 3$ kpc, 
$M_{\rm H\,{\sc i}}
\la 10^5 M_{\odot}$, and $M_{tot} \sim 10^7 M_{\odot}$.  These galaxies show 
very little central concentration and are dominated by old and intermediate-age
stellar populations.  Only three (Sgr, For, And\,I) contain globular 
clusters.  With the exception of two isolated dSphs (Tuc and Cet) all known 
dSphs are close neighbors of M31 or the Milky Way.  DSphs are gas-poor
systems.
Sensitive searches for 
H\,{\sc i} in dSphs yielded only low upper limits, but recent studies detected
extended H\,{\sc i} clouds in the surroundings of some dSphs that may be 
associated with them judging from the similarity of their radial velocities 
(Carignan et al.\ 1998, Blitz \& Robishaw 2000).    
\end{itemize} 

A few dwarf galaxies (Phe, LGS\,3) are classified as ``transition-type'' 
objects and may
be evolving from low-mass dIrrs to dSphs.  These dIrr/dSph
galaxies are found at distances of 250 kpc $< D_{\rm Spiral} < 450$ kpc.
The Local Group does not contain blue compact dwarf galaxies, dwarf spirals,
or massive ellipticals.

\section{The Local Group spirals}

The Local Group spirals have the most complex and varied star formation
histories of all Local Group galaxies.  Different subpopulations can be
distinguished by their ages, metallicities, and kinematics.  The oldest
populations are found in the halos and thick disk components.  Extremely
metal-poor ([Fe/H] $< -3$ dex) halo stars are tracers of the earliest star 
formation events (Ryan et al.\ 1996), but it is difficult to derive ages
for them.  

The earliest significant star formation episodes in the Galactic thick
disk appear to have occurred 13 Gyr ago, while the thin disk began to
experience multiple bursts of star formation $\sim 9$ Gyr ago
(Rocha-Pinto et al.\ 2000).  The 
metallicity in the thin disk depends more strongly on 
Galactocentric radius than on age and shows a large spread at any
position and age (Edvardsson et al.\ 1993).  

While halos may have largely formed through accretion of metal-poor
Searle \& Zinn (1978) fragments, bulges also host metal-rich old 
populations (mean metallicity of the Galactic bulge: $-0.25$ dex;
Minniti et al.\ 1995), 
indicating that they experienced early and fast enrichment.
M31 appears to have undergone
rapid enrichment as a whole, whereas M33 shows a pronounced radial
abundance gradient.  The mean metallicity of M31's halo is $-1$ to $-1.2$
dex, more metal-rich than the halo of the Milky Way ($\sim -1.4$ dex) and
of M33 ($\sim -1.6$ dex).  While M31's bulge emits $\sim 30$\% of the 
visible light of this galaxy, M33 lacks a bulge.  

M31's total number of 
globular clusters may be as high as $\sim 600$.  The Milky Way contains
$\sim 160$ globulars, and in the smaller M33 54 globulars are currently
known (see Grebel 2000b for a review of star clusters in the Local Group).
Main-sequence photometry of Galactic globular clusters suggests a range
of ages spanning more than 3 Gyr.  We lack such detailed information for
M31's and M33's globulars, but the blue horizontal branch (HB) morphology 
observed in some of them may suggest similar ages as for the Milky Way
globulars.  On the other hand, the red HBs of M33's
globulars may indicate that star formation was delayed by a few Gyr
(Sarajedini et al.\ 1998).  
 
The spiral arms in all three galaxies contain numerous OB associations
and young star clusters.  The UV line strengths of massive OB stars 
suggest that the young population of M31 is comparable to that of
the Milky Way, whereas M33 resembles the Large Magellanic Clouds (Bianchi, 
Hutchings, \& Massey 1996).  Present-day star-forming regions in the Milky Way
range from very extended associations to compact starburst clusters such
as the central cluster of NGC 3603 and the clusters Quintuplet
and Arches near the Galactic center.  M31's current star-forming activity 
is low.  The increase
in cluster formation in M33 over the past 10 -- 100 Myr may be correlated
with gas inflow into M33's center (Chandar, Bianchi, \& Ford 1999).   

Warps in the stellar and H\,{\sc i} disks of the Milky Way and M31 
may have been caused by tidal interactions with the Magellanic Clouds
and M32, respectively.  The Milky Way disk may also have been significantly
distorted by interacting with the currently merging Sagittarius 
dwarf galaxy (Ibata \& Razoumov 1998).
M33's stellar and H\,{\sc i} disks are tilted
with respect to each other, but no nearby companion is known that might be 
responsible. 

\section{Star formation histories of Local Group dwarf galaxies}

The star formation histories of dwarf galaxies in the Local Group vary
widely.  No two galaxies are alike; not even within the same morphological
type.  The reasons for this diversity are not understood.  It seems
that both galaxy mass and environment play important roles in 
the evolution of these low-mass objects.  

\subsection{Methods and limitations}

Star formation histories of resolved dwarf galaxies
are commonly derived through photometric techniques.
The most widely used method consists of sophisticated modelling of the 
observed color-magnitude
diagrams (CMDs) through synthetic CMDs taking into account
photometric errors, seeing, and crowding effects.  For a recent review
of procedures and techniques see Aparicio (1999). 
The methods are limited by the quality of the observations and 
by how closely theoretical evolutionary models reproduce observational
features.  For instance, Olsen (1999) notes that old red giant branches
of evolutionary models may fit the observations poorly, which can lead
to an underestimation of the contribution of the old population.  
Free parameters in modelling include 
the adopted initial mass function slope and the binary fraction.  

Additional constraints can be imposed by using special types of stars
as tracers of certain evolutionary phases.  For instance, the presence
of HB stars and RR Lyrae variables is a reliable indicator
of an old population even when sufficiently deep main-sequence photometry
is lacking.  It is important to keep in mind that the age resolution
that can be obtained is not linear and decreases strongly for older
populations.  Whereas young populations with short-lived, luminous
massive stars can be accurately age-dated to within a few million years,
the accuracy for the oldest, long-lived evolutionary phases
is of the order of a few billion years.  Relative ages of resolved
old populations with high-quality, deep main-sequence photometry,
on the other hand, can be established with a resolution of a Gyr or less
through direct comparison with CMDs of
Galactic globular clusters.  In the following, ``young''
refers to populations with ages $< 1$ Gyr, ``intermediate-age'' denotes
the age range from 1 Gyr to 10 Gyr, and ``old'' stands for ages $> 10$ Gyr.

Owing to the availability of 10-m class telescopes,
spectroscopic measurements of stellar abundances are now feasible for 
individual supergiants and the brightest red giants in galaxies as distant as  
the M31 subgroup.  Together with emission-line spectroscopy of H\,{\sc ii}
regions, these data help to constrain the metallicity and metallicity
spread in certain evolutionary phases.  Still, accurate metallicity 
information as a function of time is lacking for almost all galaxies.  

The increasing amount of data on internal kinematics and dwarf galaxy 
proper motions
are beginning to constrain their dynamical history.  Unfortunately
accurate orbital data are not yet available for almost all of the Local
Group galaxies, making it difficult to evaluate the suggested
impact of environmental effects and interactions discussed later.  

\subsection{Old populations}

A common property of all Local Group dwarfs studied in detail is the
existence of an old population, whose presence can be inferred either from
HB stars and/or from photometry reaching below the oldest
main-sequence turnoff.  Old populations may be difficult to detect in
the central portions of galaxies with significant intermediate-age or 
young populations, as the location of these stars in a CMD 
may obscure an old HB.  Also, coverage of only 
a small field of view may be insufficient to reliably detect a sparsely
populated HB (compare the findings of Gallart et al.\
1999 and Held et al.\ 2000 for Leo\,I).  {\it Age dating} of the oldest 
populations is reliably possible only where high-quality photometry 
well below the oldest main-sequence turnoff exists; a challenge 
for present-day telescopes already for galaxies at the distance of M31.
Definite statements 
about the existence of an old population are possible only where the 
photometry reaches at least the HB; feasible 
in principle with 
present-day telescopes out to distances $\approx 3$ Mpc. 
 
Deep main-sequence photometry based largely on {\em Hubble Space
Telescope} data revealed that the ages of the oldest populations 
in the LMC (Holtzman et al.\ 1999), Sagittarius (Layden \& Sarajedini 2000), 
Draco, Ursa Minor (Feltzing, Gilmore, \& Wyse 1999), Sculptor 
(Monkiewicz et al.\ 1999), Carina
(Mighell 1997), Fornax (Buonanno et al.\ 1998),
and Leo\,II (Mighell \& Rich 1996) are as old as the oldest Galactic 
globular clusters and
bulge populations.  Thus all of these galaxies share a common epoch
of early star formation.  Similarly old ages were inferred from the
existence of blue HBs in Sextans (Harbeck et al.\
2000), Leo\,I (Held et al.\ 2000), Phoenix (Smith, Holtzman, \&
Grillmair 2000), IC\,1613 (Cole et al.\ 1999), Cetus (Tolstoy et al.\ 2000), 
And\,I (Da Costa et al.\ 1996), And\,II (Da Costa et al.\ 2000), 
NGC\,185 (Geisler et al.\ 1999), NGC\,147 (Han et al.\ 1997), 
Tucana (Lavery et al.\ 1996), M31 (Ajhar et al.\ 1996),
potentially in M32 (Brown et al.\ 2000),
and spectroscopically for one of NGC 6822's globular clusters
(Cohen \& Blakeslee 1998).
Assuming that age is the second parameter determining HB 
morphology the apparent lack of a {\em blue} HB 
in M33 globular clusters (Sarajedini et al.\ 1998) and in the field 
populations of WLM (Dolphin 2000), Leo\,A (Tolstoy et al.\ 1998), 
DDO\,210 (Tolstoy et al.\ 2000) and the Small Magellanic Cloud (SMC)
may be interpreted as evidence for delayed
formation of the majority of the old population in these galaxies.  
Furthermore, the oldest globular cluster in the SMC, NGC 121, is a few 
Gyr younger than the oldest Galactic globulars (Shara et al.\ 1998) 
A complete lack 
of an old population has so far not been established in any Local Group
galaxy.

\subsection{Spatial variations of stellar populations}

Not surprisingly properties such as gas and stellar content, age structure,
metallicity distribution, density, and scale height vary as a function of 
position within a galaxy.  Spatial variations in the distribution of stellar 
populations of different ages are found in all types of galaxies, underlining
the importance of large-area coverage when trying to determine 
the star formation history of a galaxy.  

The oldest populations turn out to be spatially most extended. 
Spiral galaxies in the Local Group show pronounced population
differences between disk, halo, and more 
intricate spatially and kinematically distinct subdivisions.   
In massive irregulars such as the LMC spatial variations are traced by, e.g., 
multiple distinct regions of concurrent star formation.  
These regions can remain active for several 100 Myr, are found throughout the
main body of these galaxies, and can migrate.

In low-mass dIrrs and several dSphs the most recent star formation events
are usually centrally concentrated.  A 
radial age gradient may be accompanied by a radial metallicity gradient,
indicating that not only gas but also metals were retained over an extended
period of time.  Occasionally evidence for shell-like 
propagation of star formation from the central to adjacent regions is found.
DSphs that are predominantly old tend to exhibit radial
gradients in their HB morphology such that the ratio of red
to blue HB stars decreases towards the outer parts of the
dwarfs.  
If such second-parameter variations are caused by age then this
would indicate star formation persisted over a longer period of time in the
centers of these ancient galaxies. 

\subsection{Differences in gas content}

The H\,{\sc i} in dIrrs is generally more extended than the oldest
stellar populations and shows a clumpy distribution.  Gas and stars
in a number of low-mass dIrrs exhibit distinct spatial distributions
and different kinematic properties.  Shell-like 
structures, central H\,{\sc i} holes, or off-centered gas may be driven by
recent star formation episodes (Young \& Lo 1996; 1997a,b).  H\,{\sc i} shells,
however, do not always expand, which may argue against their formation through 
propagating star formation (Points et al.\ 1999, de Blok \&
Walter 2000).   

Ongoing gas accretion appears to be feeding the starburst in the dIrr IC 10 
(Wilcots \& Miller 1998).  An infalling or interacting H\,{\sc i}
complex is observed in the dIrr NGC\,6822 (de Blok \& Walter 2000).

DEs in the Local Group contain low amounts of gas
(a few $10^5 M_{\odot}$; Sage, Welch, \& Mitchell 1998) or none (NGC\,147).  
The apparent lack of gas in dSphs (e.g., Young 2000)
continues to be hard to understand, in particular when considering 
that some dSphs show evidence for recent (Fornax: $\sim 200$ Myr, Grebel
\& Stetson 1999) or pronounced intermediate-age star formation episodes
(e.g., Carina: 3 Gyr; Hurley-Keller, Mateo, \& Nemec 1998; Leo\,I: 2 Gyr;
Gallart et al.\ 1999).
Gas concentrated in two extended lobes along the direction of motion of
the Sculptor was detected beyond the tidal radius of this galaxy (Carignan
et al.\ 1998).  This gas may be moving inwards or away from Sculptor.
Its amount is consistent with the expected mass loss from red
giants, though that does not explain its location along the probable
orbital direction of Sculptor.
Blitz \& Robishaw (2000) suggested the existence
of similar gas concentrations with matching radial velocities in the 
surroundings of several other dSphs.  
Simulations by Mac Low \& Ferrara (1999) suggest that {\it total} gas loss
through star formation events can only occur in galaxies with masses of less
than a few $10^{-6}$ M$_{\odot}$.  Blitz \& Robishaw discuss tidal effects
as the most likely agent for the displacement of the gas.
However, the absence of gas in Cetus and Tucana, two isolated dSphs in the
Local Group, requires a different mechanism.

\subsection{Star formation histories}

Local Group dwarf galaxies vary widely in their star formation histories,
chemical enrichment, and age distribution; even within the same morphological
type.  Despite their individual differences, however, they tend to follow
common global relations between, e.g., mean metallicity, absolute magnitude,
and central surface brightness.  Galaxy mass as well as external effects
such as tides appear to play major roles in their evolution.  

Sufficiently massive irregulars and dIrrs exhibit continuous
star formation at a variable rate.  They can continue to
form stars over a Hubble time and undergo gradual enrichment.  
Galaxies such as the LMC (Holtzman et al.\ 1999, Olsen 1999),
SMC, and WLM (Dolphin 2000) have formed stars continuously and experienced
considerable chemical enrichment spanning more than 1 dex in [Fe/H].  
Their star formation {\it rate}, on the other hand, varied and shows
long periods of low activity.  Interestingly, in the LMC cluster and
field star formation activity show little correlation. 

Low-mass
dIrrs and dSphs often show continuous star formation rates 
with {\it decreasing} star formation rates.  They typically show dominant old
(or intermediate-age) populations with little or no recent activity.  A
similar evolution appears to have occurred in dEs.  DSph
companions of the Milky Way tend to have increased fractions of 
intermediate-age populations with increasing Galactocentric distance,
indicating that external effects such as tidal or ram pressure stripping
may have affected their star formation history (e.g., van den Bergh 1994).  
The two closest dSphs
to the Milky Way (other than the currently merging Sagittarius dSph)
are Draco and Ursa Minor, which are dominated by ancient 
populations and are also the least massive dSphs known -- possibly due
to the early influence of Galactic tides, though present-day positions
may not reflect early Galactocentric distances, and 
reliable orbital information is lacking. 

The Local Group dwarf galaxy to show the most extreme case of 
{\it episodic} star formation
with Gyr-long periods of quiescence and distinct, well-defined
subgiant branches is Carina (Smecker-Hane et al.\
1994, Hurley-Keller et al.\
1998).  It is unclear what caused the interruption
and subsequent onset of star formation after the long gaps.  Also,
the apparent lack of chemical enrichment during these star formation
episodes is surprising. 

\subsection{Potential evolutionary transitions}

Fornax is the second most luminous dSph galaxy in the Local Group.
The young age of its youngest measurable population ($\sim 200$ Myr, Grebel
\& Stetson 1999) is astonishing considering its lack of gas.  Just a 
few hundred Myr ago Fornax would have been classified as a dIrr.  What
caused Fornax to lose all of its gas after some 13 Gyr of 
continuous, decreasing star formation is not clear.

The presence of intermediate-age populations in some of the
more distant Galactic dSphs, the possible detection of associated gas in the
surroundings of several of them, indications of substantial mass loss
discussed elsewhere in this paper, morphological segregation, common
trends in relations between their integrated properties, and the
apparent correlation between star formation histories and Galactocentric
distance all seem to support the idea that low-mass dIrrs will eventually
evolve into dSphs if their environment fosters this evolution.  DSphs
may be the natural final phase of low-mass dIrrs, and the type distinction
may be artificial.  
The six dSph companions of M31 span a similar range in distances as the 
Milky Way dSphs (Grebel \& Guhathakurta 1999).  A study of whether their
detailed star formation histories (not yet available)
show a comparable correlation with distance from
M31 would provide a valuable test of the suggested impact of environment.  

The mass (traced by the luminosity) of a dwarf galaxy plays a major
role in its evolution as indicated by the good correlation between luminosity
and mean metallicity (e.g., Caldwell 1999).  The observed lack of 
rotation in dSphs requires that its hypothesized low-mass dIrr progenitor
must have gotten rid
of its angular momentum, which may occur through substantial mass loss.  
However, this scenario does not account for the existence of isolated
dSphs such as Tucana.  Alternatively, the progenitor may have had very
little rotation to begin with. 
Either way, the subsequent fading must have been low since 
otherwise dIrrs and dSphs would not follow such a fairly well-defined
common relation.  Several authors (e.g., Mateo 1998) suggested that
the luminosity-metallicity relation is instead bimodal with separate
loci for dIrrs and dSphs in the sense that at a given luminosity
a dIrr tends to be more metal-poor than a dSph, excluding evolutionary
transitions.
Hunter, Hunsberger, \& Roye (2000) go a step further and suggest that a number 
of Local Group dIrrs
might have formed as ancient tidal dwarfs that lack dark matter, are
essentially non-rotating, and contribute to the increased scatter in
the absolute magnitude--mean metallicity relationship for $M_B<-15$ mag. 

\section{Dark matter}

Dark matter is a significant component of many Local Group 
galaxies.  Spiral galaxies exhibit H\,{\sc i} rotation curves that
become approximately flat at large radii and that extend 2--3 times 
beyond the optically visible galaxy.  Global mass-to-light ratios (M/L)
inferred from rotation curves of spirals are typically $\le 10$ 
M$_{\odot}$/L$_{\odot}$ for the visible regions ($\sim 1 - 3$ 
M$_{\odot}$/L$_{\odot}$ in disks, $\sim 10 - 20$ 
M$_{\odot}$/L$_{\odot}$ in bulges), while the dark matter in halos
seems to significantly exceed these values (Longair 1998).  
This motivates efforts to determine the nature of the dark matter
through microlensing in the Galactic halo and toward the Galactic
bulge as detailed elsewhere in this volume, and through pixel microlensing
of stars in the disk of M31 by dark massive objects in M31's halo (Crotts
1992).

In gas-rich dwarfs the presence of dark matter is inferred as well from
H\,{\sc i} rotation curves.
Some of the less massive dIrrs are rotationally supported
only in their centers, while the majority of dSphs studied so far does not
show evidence for rotation at all.  Chaotic gas motions dominate in low-mass
dIrrs, and the H\,{\sc i} column density distribution
is poorly correlated with the stellar distribution (Lo, Sargent, \& Young
1993).
In the dE NGC\,205, which is tidally interacting with M31, 
integrated light measurements revealed that the stellar
component is essentially non-rotating though the H\,{\sc i} shows 
significant angular momentum (Welch, Sage, \& Mitchell 1998).  
In gas-deficient dSphs kinematic information is based entirely on
stars.  Most dSphs show no rotation.
Their velocity dispersions are typically $\ge 7$ km s$^{-1}$.   
Assuming virial equilibrium 
velocity dispersions and rotation curves can be translated into virial masses. 
The derived total M/L ratios 
of Local Group dwarf galaxies present an inhomogeneous picture ranging 
from $\sim 1$ to $\sim 80$ (see compilation by Mateo 1998).  

Compact high-velocity clouds (CHVCs) are a subset of high-velocity H\,{\sc i}
clouds with angular sizes of only about 1 degree on the sky.  They show
infall motion with respect to the barycenter of the Local Group.
Preliminary estimates 
place them at distances of 0.5 to 1 Mpc in contrast to the extended
nearby high-velocity-cloud complexes (Braun \& Burton 1999).  Their 
rotation curves
imply high dark-to-H\,{\sc i} ratios of 10--50 if distances of 0.7 Mpc
are assumed, and masses of $10^7$ M$_{\odot}$ (Braun \& Burton 2000).  
CHVCs may be a significant source of dark matter and may
represent pure H\,{\sc i}/dark-matter halos prior to star formation.
We are currently carrying out an optical wide-field survey to establish
whether they also contain a low-luminosity, low-density stellar
component, which
would imply the discovery of a new, very dark type of galaxy,
help to refine CHVC distances and allow detailed studies of their
stellar populations.

\subsection{Dwarf spheroidal galaxies and dark matter}

Galactic dSphs are of particular interest in efforts to elucidate the
nature of dark matter since they may be dark-matter-dominated and can  
be studied in great detail due to their proximity.
{From} an analysis of the kinematic properties of Draco and Ursa Minor
Gerhard \& Spergel (1992a) exclude  fermionic light particles (neutrinos)
as dark matter suspects because 
phase-space limits would then require unreasonably large core radii and 
masses for these two galaxies.

The initial measurements of velocity dispersions in dSphs were criticized 
for including luminous AGB stars and Carbon stars,
whose radial velocities may reflect atmospheric motions, and for neglecting
the impact of binaries (see Olszewski 1998 for details).   
Subsequent studies concentrated on somewhat fainter stars along the upper
RGB, carried out extensive simulations to assess the impact of binaries
(Hargreaves, Gilmore, \& Annan 1996; Olszewski, Pryor, \& 
Armandroff 1996), obtained multi-epoch observations (e.g., Olszewski, 
Aaronson, \& Hill 1995),
and increased the number of red giants with measured radial velocities to
more than 90 in some cases (Armandroff, Olszewski, \& Pryor 1995).  
These studies established that the large velocity dispersions in dSphs
are not due to the previously mentioned
observational biases.  Kleyna et al.\ (1999) show that the
currently available measurements for the two best-studied dSphs, Draco and
Ursa Minor, are not yet sufficient to distinguish between models where
mass follows light (constant M/L throughout the dSph)
or extended dark halo models when interpreting the 
velocity dispersions as high M/L ratios due to large dark matter content.
Mateo (1998) and Mateo et al.\ (1998) argue that the relation between
total M/L and $V$-band luminosity for dSphs can be approximated well
when adopting 
a stellar M/L of 1.5 (similar to globular clusters) and an extended dark
halo with a mass of $2\cdot10^{7}$ M$_{\odot}$, suggesting fairly uniform
properties for the dark halos of dSphs.

Luminosity functions (LFs) of old stellar systems can provide further 
constraints on the nature of dark matter.  The main-sequence LFs of old  
field populations in the Galactic bulge (Holtzman et al.\ 1998),
LMC and SMC (Holtzman et al.\ 1999), Draco (Grillmair et al.\ 1998),
and Ursa Minor (Feltzing et al. 1999) are
in excellent agreement with the solar neighborhood IMF and 
LFs of globular clusters that did not suffer mass segregation. 
Since globular clusters are not known to contain dark matter, one would expect
to find differences in the LF of dark-matter-rich populations if low-mass
objects down to 0.45 M$_{\odot}$ were important contributors to the
baryonic dark matter content.  Furthermore, these studies demonstrate
that the LF in objects with a wide range of M/L ratios does not differ much.
The possible contribution of white dwarfs (or lack thereof)
is discussed elsewhere in these proceedings.

\subsection{Tidal effects rather than dark matter?}

Instead of a smooth surface density profile that one might expect
from a relaxed population, Ursa Minor shows statistically significant 
stellar density variations (Kleyna et al.\ 1998).
Fornax's four ancient globular clusters
are located at distances larger than the galaxy's core radius.  Dynamical
friction should have lead to orbital decay in only a few Gyr (much
less time than the globular clusters' lifetimes) and have turned Fornax
into a nucleated dSph.  Simulations by Oh, Lin, \& Richer (2000)
suggest that the best mechanism to have prevented this evolution is
significant mass loss through Galactic tidal perturbation and the resulting
decrease in the satellite galaxy's gravitational potential, which
may have increased the clusters' orbital semimajor axes and efficiently
counteracted the spiralling-in through dynamical friction.
The detection of a possible extended population of extratidal stars 
around the dSph Carina
might imply that this galaxy has now been reduced to a mere 1\% of its
initial mass (Majewski et al.\ 2000).  If such significant tidal disruption
is indeed real and widespread then the present-day stellar content of nearby
dSphs cannot easily be used to
derive evolutionary histories over a Hubble time.  Furthermore, extended
extratidal stars are not expected if the galaxy is dark-matter dominated
(Moore 1996).

Additional indications in favor of the impact of galactic tides come from the 
structural parameters of dSphs (Irwin \& Hatzidimitriou 1995), which 
seem to imply tidal disruption in several cases.
Furthermore, substantial tidal disruption by the Milky Way
is evidenced by the Magellanic Clouds and
Magellanic stream, by the Sagittarius dSph, and by Galactic
globular clusters (Gnedin \& Ostriker 1997; Grillmair et al.\ 1995;
Leon, Meylan, \& Combes 2000).
Tracer features include gaseous and stellar tidal tails (Putman et al.\
1998; Majewski et al.\ 1999; Odenkirchen et al.\ 2000).  
The conversion of velocity dispersions into 
M/L ratios and dark matter fractions assumes 
virial equilibrium, a condition that is violated in the case of 
severe tidal disruption.  

Tidal heating due to resonant orbital coupling between the time-dependent
Galactic gravitational field and the internal oscillation time scales of
dSphs (Kuhn \& Miller 1989; Kuhn 1993; Kuhn, Smith, \& Hawley 1996) may 
inflate the dSphs' velocity dispersions, but see Pryor (1996) for arguments
against the efficiency of this mechanism.   

As shown by Piatek \& Pryor (1995)
tides can, but need not inflate the global M/L/ratio to values. 
Indeed, in a galaxy suffering tidal disruption the velocity dispersion
can be sustained at its virial equilibrium value, and the central
density is maintained even after substantial mass loss (Oh, Lin, \&
Aarseth 1995).  Pryor (1996) noted that a velocity gradient across
a galaxy that is larger than the 
velocity dispersion is the clearest signature of tidal disruption,
but such a gradient is not obvious in the Galactic dSphs.  

Kroupa (1997) and Klessen \& Kroupa (1998) proposed that stellar tidal
tails may look like dSphs when seen along the line of sight, an orientation
that follows naturally from their N-body simulations.  The ordered motions
in the tidal remnants would appear as increased velocity dispersion since
they occur along the line of sight.  These models can roughly reproduce the 
observed correlations between central surface brightness, absolute
magnitude, and M/L.  The predicted line-of-sight extension of the dSphs
can be tested, in principle, through accurate measurements of the 
apparent width of their 
HBs.  The predicted high orbital eccentricity, a consequence
of the required radial orbits in the model, could be checked  
through accurate proper motion measurements with astrometric 
satellite missions such as {\em SIM} and {\em GAIA}.
The tidal remnants may be leftovers from earlier
mergers as suggested by the observation that the Galactic dSph
galaxies appear to be located near at least two polar planes or great circles
(the Magellanic Stream and the Fornax--Leo--Sculptor
Stream; e.g., Kunkel \& Demers 1996; Kunkel 1979; Lynden-Bell 1982;
Majewski 1994).  Such tidal remnants would not likely contain dark
matter.  The observed ages and abundances of galaxies potentially associated
with ``streams'' constrain the time at which the break-up of a more massive
parent could have occurred.  This event must have happened
very early on when the parent had not yet experienced significant enrichment. 
Siegel \& Majewski (2000) suggest that galaxies potentially belonging to
a stream may have originated from a common $-2.3$ dex progenitor and
subsequently followed their own evolution. 

The impact of Galactic tides remains a valid alternative to large amounts of
dark matter in nearby dSphs.  The determination of velocity dispersions of
distant or even isolated dSphs, which are unlikely to be subject to tidal
effects, is an important test of whether high M/L ratios in dSphs are 
largely caused by environmental effects (see, e.g., the discussion in
Bellazzini, Fusi Pecci, \& Ferraro 1996).  Stellar velocity dispersions 
indicative of high M/L were found in the most distant ($\sim 270$ kpc)
potential Milky
Way dSph companion Leo\,I (Mateo et al.\ 1998) and in the outlying 
($\sim 280$ kpc) M31 transition-type satellite LGS\,3 (Cook et al.\
1999), but measurements of truly isolated Local Group dSphs such as Tucana
and Cetus are still lacking. 

\subsection{Modified Newtonian dynamics}

Modified Newtonian dynamics (MOND, Milgrom 1983a,b), which alters
Newton's second law at low accelerations by introducing a multiplicative
acceleration constant of $1.2 \cdot 10^{-8}$ cm s$^{-1}$, results in 
M/L ratios that do not require the presence of dark matter.  While many
attempts have been made to disprove MOND (e.g., Gerhard \& Spergel 1992b), 
none of the presently existing 
measurements has been able to unambiguously refute MOND for either disk
galaxies (van den Bosch \& Dalcanton 2000) or dwarfs (e.g, Milgrom
1994; 1995; C\^ot\'e
et al.\ 1999).  MOND remains a possible alternative to dark matter.

\subsection{Implications for microlensing}

Microlensing surveys are concentrating on the Galactic bulge, the Galactic
halo through monitoring of sight lines toward the Magellanic Clouds, and
on M31 through pixel lensing.  All of these surveys 
concentrate on fields with high-density, luminous background populations.  
Results and 
constraints on dark matter from these surveys are discussed elsewhere in
this volume.

The remaining Local Group
dwarf galaxies are less well suited for classical microlensing studies. 
Advantages of using other dwarfs are that one can probe 
additional lines of sight and can take advantage of the large optical
depth to microlensing since the sources are outside of the Milky Way
halo.  Also, in nearby dSphs crowding won't be much of a problem.  
However, the efficiency of such studies would be drastically reduced as 
compared to the ongoing studies since the targets are faint and stellar
densities are low.  This requires not only longer exposure times or larger
telescopes but also implies much longer time scales before a significant
number of events can be observed.  

As discussed earlier the high velocity dispersions in low-mass dwarfs
may arise from large amounts of dark matter.  This increases the
possibility that one may observe self-lensing when monitoring dwarfs
rather than events in the Galactic halo, an effect that is negligible
when turning to distant Galactic globular clusters instead (Gyuk \&
Holder 1998).  As always, variable stars
may act as contaminants.  
Future large survey telescopes 
such as the proposed {\em Dark Matter Telescope} (Tyson, Wittman, \&
Angel 2000) can provide routinely deep exposures of nearby Local Group
dwarf galaxies once per night as a regular 
by-product of their search for cosmological weak lensing. 

\section{Summary}

The Local Group, an ensemble of 35 galaxies most of which are dwarf 
companions of either M31 or the Milky Way, contains galaxies with a 
wide variety of masses, luminosities, star formation histories, and
chemical and kinematic properties.  No two galaxies in the Local Group
experienced the same star formation history even within the same
morphological type.  Star formation episodes
vary in length and times ranging from continuous star formation with
variable star formation rates to gradually declining rates and episodic
star formation, accompanied by either gradual chemical enrichment or 
almost no enrichment at all.  Old populations are a common property of
all Local Group galaxies studied in detail so far, though not all appear
to share a common epoch of the earliest measurable star formation.    
Spatial variations in ages and abundances are observed in most Local
Group galaxies ranging from widely scattered active regions in
high-mass galaxies to centrally concentrated younger star formation 
episodes in low-mass dwarfs.
Both galaxy mass and galaxy environment appear to have a 
major impact on galaxy evolution.  Interactions such as ram pressure
and tidal stripping seem to influence the evolution of less massive
galaxies, contributing to the observed morphological segregation.  
Rotation curves and stellar velocity dispersions indicate the presence
of dark matter in the majority of Local Group galaxies, although 
alternative explanations cannot be ruled out.  The properties of the 
stellar populations in these galaxies as well as orbital and kinematic
information can impose constraints on the nature and ubiquity of 
dark matter.  Owing to faintness, low
stellar density and hence low event probability as well as 
likeliness of self-lensing, Local Group dwarf galaxies other than the
Magellanic Clouds are poorly suited for classical microlensing surveys
but might become of interest for future large telescopes that routinely
monitor a major fraction of the sky on a nightly basis.

\acknowledgments

This work was supported by NASA through grant HF-01108.01-98A from the
Space Telescope Science Institute, which is operated by the Association of
Universities for Research in Astronomy, Inc., under NASA contract NAS5-26555.
I gratefully acknowledge support from the organizers who covered my local
expenses.  Last but not least I like to thank the editors for their patience
while this manuscript was finished.

\end{document}